\documentclass[twocolumn,english,prx,superscriptaddress]{revtex4-1}
\usepackage[utf8]{inputenc}
\usepackage{xcolor}
\usepackage{babel}
\usepackage{bm}
\usepackage{graphicx}
\usepackage{amsmath}
\usepackage{amssymb}
\usepackage{empheq}
\usepackage{times}

\usepackage[colorlinks=true,citecolor=blue]{hyperref}
\begin{document}

\title{Time optimal quantum state transfer in a fully-connected quantum computer}

\author{Casey Jameson}
\email{cwjameson@mines.edu}
\affiliation{Department of Physics, Colorado School of Mines, Golden, Colorado 80401, USA}

\author{Bora Basyildiz}
\affiliation{Department of Computer Science, Colorado School of Mines, Golden, Colorado 80401, USA}

\author{Daniel Moore}
\affiliation{Department of Physics, Colorado School of Mines, Golden, Colorado 80401, USA}

\author{Kyle Clark}
\affiliation{Department of Physics, Colorado School of Mines, Golden, Colorado 80401, USA}

\author{Zhexuan Gong}
\email{gong@mines.edu}
\affiliation{Department of Physics, Colorado School of Mines, Golden, Colorado 80401, USA}
\affiliation{National Institute of Standards and Technology, Boulder, Colorado 80305, USA}

\begin{abstract}
      
The speed limit of quantum state transfer (QST) in a system of interacting particles is not only important for quantum information processing, but also directly linked to Lieb-Robinson-type bounds that are crucial for understanding various aspects of quantum many-body physics. For strongly long-range interacting systems such as a fully-connected quantum computer, such a speed limit is still unknown. Here we develop a new Quantum Brachistochrone method that can incorporate inequality constraints on the Hamiltonian. This method allows us to prove an exactly tight bound on the speed of QST on a subclass of Hamiltonians experimentally realizable by a fully-connected quantum computer.

\end{abstract}

\maketitle

As investment into quantum computing platforms continues to grow and the scale of such platforms rapidly expands \cite{Preskill_2018}, transferring a quantum state between two distant qubits becomes an important problem. If the physical interactions among the qubits are short-ranged, then the speed of information propagation is always finite according to the Lieb-Robinson bound \cite{LiebOriginal}. As a result, the time for performing quantum state transfer (QST) within such systems has to be at least proportional to the distance between the two qubits. Recently however, long-range interacting quantum systems have become increasingly popular for both quantum computing and quantum simulation, with these systems including trapped ions \cite{MonroeSimulation}, Rydberg atoms \cite{Saffman2010}, polar molecules \cite{Yan2013}, defect centers in solids \cite{Yao2012}, atoms coupled to cavities and photonic crystals \cite{Douglas2015}.
The physical interactions in these systems typically decay as a power law $1/r^{\alpha}$ with $r$ being the inter-particle distance. For sufficiently small $\alpha$, these long-range interactions can be utilized to provide speedups for quantum information processing \cite{Linke2017,EldredgeStateTransfer,TranStateTransfer, FossFeigSpinSqueeze} or realize novel quantum many-body dynamics and phases \cite{Richerme2014,Jurcevic2014,Smith2016,Zhang2017a,Neyenhuis2017,Feng2022,Chen2022}. However, the exact degree of possible speed up for performing QST or other important quantum information processing tasks is unknown \cite{TranStateTransfer}.

More concretely, with the recent development of optimal Lieb-Robinson-type bounds for a wide-range of long-range interacting systems \cite{Hastings2006a,Gong2014,FossFeig2015,Kuwahara2020,Chen2019,TranHeirarchy,Tran2021}, it has been shown that the speed of QST remains finite as long as $\alpha>2D+1$, where $D$ is the dimension of the system. However, for systems where the interaction decays with $\alpha<D$, no such optimal Lieb-Robinson-type bound has been found. Nonetheless, these systems are important as they could provide a dramatic speedup in quantum information processing. For example, a fully-connected quantum computer \cite{Linke2017} (where $\alpha=0$) could allow exponentially faster creation of topologically ordered states when compared to quantum computers with short-range interacting qubits \cite{EldredgeStateTransfer}. With respect to the QST, the best known Lieb-Robinson-type bound \cite{GuoScrambling} predicts a time of at least $O(\log(N)/N)$ for $\alpha=0$, but the fastest known QST protocol in this case requires a minimum time of $O(1/\sqrt{N})$ \cite{GuoScrambling}.

While it remains challenging to find an optimal Lieb-Robinson-type bound for generic strongly long-range interacting systems, here we take a different but novel approach to bound the speed limit of QST, using the framework of the Quantum Brachistochrone (QB) \cite{Carlini2006,CarliniAction,CarliniUnitary,RussellZermelo,WangGeodesic,KhanejaSpinControl}. This approach requires us to study only a subclass of Hamiltonians relevant for a fully-connected quantum computer but allows us to obtain an exactly tight bound for the speed of QST. To the best of our knowledge, no such bound has been previously obtained for long-range interacting many-body systems. Our work, therefore, bridges an important gap between the pursuit of optimal Lieb-Robinson-type bounds and the study of quantum speed limits, as the former often only predicts the correct scaling in the large system size limit \cite{LiebOriginal,Kuwahara2020,Chen2019,TranHeirarchy,Tran2021} while the latter is precise but is usually only obtained for small quantum systems \cite{FreyQSL,MLOriginal,MandelstamOriginal,Carlini2006,CarliniAction,CarliniUnitary,RussellZermelo,WangGeodesic,KhanejaSpinControl}. In addition, our work extends the existing study of QST in spin systems \cite{BoseQST2003,Christandl2004,Christandl2005,Godsil2012,Yao2011,Yao2013,Kempton2016} that is largely limited to time-independent Hamiltonians and thus does not predict a speed limit for general, time-dependent Hamiltonians studied here. Our work is also complementary to other optimal control methods for speeding up quantum information processing tasks such as the shortcut to adiabaticity \cite{Torrontegui2013,GueryOdelin2019,Deffner2014,Huang2018,Song2016}, which seeks to speedup adiabatic processes while preserving robustness, but does not guarantee a time optimal path in general. 

The QB method is an important tool in quantum optimal control theory that aims to find the fastest Hamiltonian to achieve a certain unitary or a particular state evolution. While the method can provide an exact speed limit for quantum information processing tasks, it is typically only used for small and simple quantum systems (such as one or two qubits) \cite{CarliniUnitary,RussellZermelo,WangGeodesic,KhanejaSpinControl}. More importantly, the application of QB typically requires the control parameters of the Hamiltonian to satisfy equality constraints. This is not suitable for studying the speed limit of QST or Lieb-Robinson-type bounds, which usually involves only an upper bound on the strengths of two-body interactions in the Hamiltonian. Therefore, a major contribution of this work is to generalize the standard QB method to incorporate inequality constraints. While there exists other methods in optimal control theory, such as those using Pontryagin's maximum principle \cite{BoscainOptControl}, that can also deal with inequality constraints \cite{WakamuraMP}, our QB based method is technically simpler and more intuitive. 

The remaining content of this paper is organized as follows: In Section \ref{SetupSection}, we introduce the particular problem we are trying to solve, i.e. finding the speed limit of QST for a subclass of Hamiltonians realizable by a fully-connected quantum computer.  In Section \ref{QBSection}, we introduce the standard QB method and our extension of it to incorporate inequality constraints. In Section \ref{TightBound}, we analytically solve the equations from our generalized QB method and obtain an exact speed limit for the QST defined in Section \ref{SetupSection}. The details of the solution are presented in the Appendix A. Finally, in Section \ref{Conclusion}, we discuss the implications of our results along with future directions our work can point toward.

\section{The quantum state transfer problem}\label{SetupSection}
In this paper, we consider the problem of (perfect) QST between two qubits in an $N$-qubit system \cite{BoseQST2003}. Specifically, the goal of QST is to transfer an arbitrary unknown state $c_0 |0\rangle + c_1|1\rangle$ from one qubit to another qubit. Without loss of generality, we may label the source qubit as our qubit $1$ and the target qubit as our qubit $N$. We may further assume that the target qubit is initially in the state $|0\rangle$ before the QST and the source qubit is in state $|0\rangle$ after the QST. This assumption can be easily made true via single-qubit gates. Finally, we will always assume all remaining qubits $\{2,3,\cdots, N-1\}\equiv \mathcal{A}$ to be in some state $|\psi_{\mathcal{A}}\rangle$ before and after the QST. With these assumptions in place, the process of QST can be expressed as:

\begin{equation}\label{QST}
(c_0 |0\rangle + c_1|1\rangle)\otimes |\psi_{\mathcal{A}}\rangle \otimes |0\rangle  \rightarrow |0\rangle \otimes |\psi_{\mathcal{A}}\rangle \otimes (c_0 |0\rangle + c_1|1\rangle)
\end{equation} 

Next, we define the class of Hamiltonians considered to achieve the QST described above. For most controllable quantum systems, the interactions among the qubits are two-body, and we can write a general form of the Hamiltonian as:
\begin{equation}\label{generalH}
H = \sum_{i\ne j}h_{ij}(t) + \sum_{i} h_i(t)
\end{equation} 
Due to the allowed time dependence, this Hamiltonian can represent any quantum circuit made of single-qubit and two-qubit gates, and can thus achieve universal quantum information processing. For most experimental systems, the two-qubit interaction strength, quantified by $\lVert h_{ij}\rVert$ where $\lVert\cdot\rVert$ denotes the operator norm of $h_{ij}$, is limited by the hardware design, while the strength of single-qubit terms $\{\lVert h_i\rVert\}$ can usually be made much larger via strong applied drives or fields. As a result, it is reasonable to assume that $\{\lVert h_i\rVert\}$ are not constrained while $\lVert h_{ij}\rVert$ is upper bounded by $J_0/r_{ij}^{\alpha}$, where $J_0$ denotes the maximum interaction strength and $\alpha\ge0$ captures how fast interactions decay in the inter-particle distance $r_{ij}$ between qubits $i$ and $j$. This assumption is made in most studies of Lieb-Robinson-type bounds \cite{Hastings2006a, Gong2014, FossFeig2015, Chen2019, Kuwahara2020, TranHeirarchy,Tran2021}.

In the Heisenberg picture, a Lieb-Robinson-type bound is typically written
\begin{equation}\label{LR}
\lVert [A_i(t),B_j(0)]\rVert \le f(t,r_{ij})
\end{equation}
where the operators $A_i$ and $B_j$ act on qubits $i$ and $j$, respectively, at $t=0$. The function $f(t,r_{ij})$ depends on the exact form of the bound and typically increases monotonically as $t$ increases.

To see why Eq.\,\eqref{LR} can lower bound the time it takes to perform the QST defined in Eq.\,\eqref{QST}, we set $B_j=\sigma_1^x$ and $A_i=\sigma_N^y$ at $t=0$, and assume qubit 1 is initially in the state $|0\rangle$. Then it is simple to see $\langle [A_i(t),B_j(0)]\rangle =2i$ for a QST assumed to occur in time $t$. As a result, we must have $f(t,r_{1N})\ge ||[A_i(t),B_j(0)]||\ge 2$, which can only be satisfied if $t>t_{\text{LR}}$ where $t_{\text{LR}}$ depends on $f(t,r_{ij})$ and $r_{1N}$. If the globally minimum time to perform QST is denoted by $t_{\text{min}}$, we call the bound Eq.\,\eqref{LR} (qualitatively) optimal when $t_{\text{LR}}\ge c t_{\text{min}}$ where the constant $c$ is independent of $r_{1N}$ and $N$. In other words, the speed limit predicted by the Lieb-Robinson-type bound is only loose by a constant factor.

Optimal Lieb-Robinson-type bounds have recently been found for a wide-range of quantum information processing tasks \cite{Chen2019,Kuwahara2020,Tran2021} when $\alpha>D$ but remain largely undiscovered for the case where $\alpha<D$. For $\alpha<D$, we often call interactions ``strongly long-ranged" \cite{GuoScrambling} since the total interaction energy per qubit diverges in $N$. In particular, the case of $\alpha=0$ corresponds to a fully-connected universal quantum computer where any two qubits can interact at equal strength. Such a quantum computer can be experimentally realized by trapped ions \cite{Linke2017}, superconducting qubits \cite{Strauch2008}, atoms coupled to cavities or waveguides \cite{Douglas2015,Park2022}, etc. In a fully-connected quantum computer, the physical distance between any two qubits is  irrelevant, and the QST time should only depend on $N$. The best known Lieb-Robinson-type bound \cite{GuoScrambling} predicts a minimum QST time $t_{\text{min}} \ge O(\log(N)/N)$. However, this bound is believed to be non-optimal as the fastest QST protocol known for $\alpha=0$ takes time $t=O(1/\sqrt{N})$.

It is worth mentioning that the fastest QST protocols known for long-range interacting systems so far are conditioned on the intermediate qubits $\{2,3,\cdots,N-1\}$ being in some specific state, usually the state $|\psi_{\mathcal{A}}\rangle=|0\rangle ^{\otimes {N-2}}$ \cite{EldredgeStateTransfer,GuoScrambling,TranHeirarchy,TranStateTransfer,Tran2021}. An unconditional QST, which is similar to a remote two-qubit gate, does not have this requirement of the intermediate qubits being in some specific state, but appears to take a much longer time than a conditional QST \cite{TranStateTransfer}. We therefore only focus on conditional QST in this paper as our goal is to 
find the fastest QST protocol for a fully-connected quantum computer. In particular, we will assume from this point that all intermediate qubits are in the $|0\rangle$ state before and after the QST. This condition can be easily realized via initial state preparation on a quantum computer.

In order to find an exact speed limit of QST, we need to further constrain the general Hamiltonian expressed in Eq.\,\eqref{generalH}. A reasonable constraint is to require the Hamiltonian to conserve the total $\sigma^z\equiv \sum_{i=1}^N \sigma_i^z$. This constraint is compatible with the QST process defined in Eq.\,\eqref{QST}, and leads the Hamiltonian in Eq.\,\eqref{generalH} to simplify to
\begin{equation}\label{SpinHam}
H(t)=\sum_{i\neq j}(J_{ij}(t) \sigma_i^{+}\sigma_j^{-}+\mathrm{h.c.}+U_{ij}(t)\sigma_i^z\sigma_j^z)+\sum_{j}B_j(t) \sigma_j^z
\end{equation}

We emphasize that this constraint is in fact natural for most quantum computing or simulation platforms where the physical interactions among qubits are flip-flop, Heisenberg, or XXZ interactions. Such a constraint is also assumed in most previous studies on QST in spin systems \cite{Christandl2004,Christandl2005,Godsil2012,Yao2011,Yao2013,Kempton2016}.
For platforms with Ising interactions, this constraint can also be approximately achieved by applying a large transverse field in the $z$ direction, effectively reducing the Ising interactions to flip-flop interactions \cite{Richerme2014,Feng2022}. We also note that the ZZ-interaction terms $\{U_{ij}(t)\sigma_i^z\sigma_j^z\}$ are not important as they play the same role as the field terms $\{B_j(t) \sigma_j^z\}$ for our QST process because we only ever have one qubit in state $|1\rangle$. Thus we will only need to constrain the flip-flop interaction strength. For a fully-connected quantum computer, the constraint we apply to Eq.\,\eqref{SpinHam} will simply be
\begin{equation}\label{constraint}
|J_{ij}(t)| \le J_0
\end{equation}

A bound on the speed of QST in this case has been derived by mapping Eq.\,\eqref{SpinHam} to non-interacting bosons/fermions, which predicts $t_{\text{min}} \ge 1/(J_0\sqrt{N-1})$ \cite{GuoScrambling}. This bound is qualitatively optimal as Ref.\,\cite{GuoScrambling} includes a QST protocol with $t=\pi/(J_0\sqrt{N-2})$ ($N\ge3$ assumed). 

\vspace{5pt}
The main results of this paper are:
\vspace{-5pt}
\begin{enumerate}
    \item For any $N\ge2$, we find a family of spin Hamiltonians of the form in Eq.\,\eqref{SpinHam} that achieves perfect QST in time
    \begin{equation}
    t=\frac{\pi}{J_0\sqrt{2N}}.
    \end{equation}
    \item We prove rigorously that this is the globally minimum QST time assuming the Hamiltonian in Eq.\,\eqref{SpinHam} is invariant under the permutations of the qubits $2,3,\cdots,N$.
\end{enumerate}

The additional assumption on permutation symmetry is natural due to the constraint in Eq.\,\eqref{constraint} and the setup of our QST. We do not believe a faster QST protocol could exist without this assumption and will provide justifications and possible ways to prove them in Section \ref{Conclusion}. This assumption is also not needed for $N=3$.

The permutation symmetry allows us to restrict the QST to within a subspace spanned by the following four basis states:
\begin{align}\label{3basis}
|\phi_0\rangle  &\equiv |0\rangle \otimes |0\rangle ^{\otimes (N-2)} \otimes |0\rangle, \\
|\phi_1\rangle  &\equiv |1\rangle \otimes |0\rangle ^{\otimes (N-2)} \otimes |0\rangle, \\
|\phi_2\rangle &\equiv |0\rangle \otimes |W_{N-2}\rangle \otimes |0\rangle, \\
|\phi_3\rangle  &\equiv |0\rangle \otimes |0\rangle ^{\otimes (N-2)} \otimes |1\rangle,
\end{align}
where $|W_{N-2}\rangle \equiv \frac{1}{\sqrt{N-2}} (|10\cdots 0\rangle + \cdots + |0\cdots 1\rangle)$ is the W-state for qubits $2,3,\cdots,N-1$. Since $|\phi_0\rangle$ is an eigenstate of the Hamiltonian in Eq.\,\eqref{SpinHam}, the non-trivial dynamics of the Hamiltonian can be captured by an effective 3-level Hamiltonian in the subspace spanned by $\{|\phi_1\rangle,|\phi_2\rangle,|\phi_3\rangle\}$:
\begin{equation}\label{HamThreeDiag}
H_{\text{eff}}(t)=
\begin{pmatrix}
\delta_1(t)& J_{1\mathcal{A}}(t)&J_{1 N}(t)\\
J^*_{1\mathcal{A}}(t)&\delta_{\mathcal{A}}(t)& J_{\mathcal{A} N}(t)\\
J^*_{1 N}(t)& J^*_{\mathcal{A}N}(t)&\delta_N(t)
\end{pmatrix},
\end{equation}
where the constraint given by Eq.\,\eqref{constraint} now becomes
\begin{equation}\label{3lvlconstraint}
|J_{1N}(t)| \le J_0, \quad |J_{1\mathcal{A}}(t)|,|J_{\mathcal{A} N}(t)|\le \sqrt{N-2} J_0.
\end{equation}
while $\delta_1(t),\delta_{\mathcal{A}}(t),\delta_N(t)$ are unconstrained. Now our goal is to find the optimal $H_{\text{eff}}$ satisfying the constraints in Eq.\,\eqref{3lvlconstraint} that takes the state $|\phi_1\rangle$ to $|\phi_3\rangle$ in minimal time. This is a much simpler problem but still highly non-trivial due to the time dependence of all matrix elements in $H_{\text{eff}}$, such that we are optimizing over infinitely many possible 3-level Hamiltonians. As it is impossible to find the analytical solution to the evolution operator of Eq.\,\eqref{HamThreeDiag}, one cannot obtain the time optimal Hamiltonian directly.

The QB method \cite{Carlini2006,CarliniAction,CarliniUnitary,RussellZermelo,WangGeodesic,KhanejaSpinControl} offers us a way to deal with this type of problem, but it does not permit inequality constraints as given in Eq.\,\eqref{3lvlconstraint}. As a result, we will need to first generalize the QB method to handle our constraints, as detailed in the next section.

\section{A Quantum Brachistochrone Method for inequality constraints}\label{QBSection}

We introduce here an extension to the QB method first developed in Ref.\,\cite{Carlini2006} that allows for the consideration of problems containing both inequality and equality constraints on the Hamiltonian. This extension is not only necessary to find the time optimal Hamiltonian of the form Eq.\,\eqref{HamThreeDiag} under the constraints in Eq.\,\eqref{3lvlconstraint}, but may also be helpful in dealing with more general Hamiltonians in Eq.\,\eqref{generalH} with bounded interaction strengths.

We note that Ref.\,\cite{WakamuraMP} has also developed an extension to the standard QB method that allows for inequality constraints, but it uses Pontraygin's maximum principle \cite{BoscainOptControl} and the formalism is quite involved. Moreover, it does not intuitively demonstrate the physical difference between a QB problem with equality and inequality constraints. Here we utilize a distinct approach based on the method of slack variables, which intuitively speaking, divides the optimization problem into two scenarios: one where the inequality constraint is simply removed and the other where the inequality constraint becomes an equality constraint. 

Formally, the method of slack variables allows us to change all inequality constraints on the Hamiltonian to equality constraints in the form of $f_j(H,s_j)=0$ for the $j^{\text{th}}$ constraint, where $s_j$ is a slack variable. For example, the constraint $|J_{1N}(t)| \le J_0$ can be rewritten as $|J_{1N}(t)|^2+|s(t)|^2 = |J_0|^2$ with a slack variable $s(t)$ added.

We can now introduce our generalized QB method, which also includes a brief review of the standard QB method. Suppose we want to find the minimal evolution time $T$ for a particular Hamiltonian $H(t)$ to generate a target unitary operator $U_0$. We can formulate this problem as an optimization problem for minimizing the functional $T=\int_0^T dt$ over the Hamiltonian $H(t)$ and the corresponding evolution operator $U(t)$ under the constraints of the Hamiltonian specified by $\{f_{j}(H,s_j)\}$ and the Schrodinger's equation $dU(t)/dt = -i H(t)U(t)$. We can then define an action integral \cite{Carlini2006,CarliniUnitary,WakamuraMP}
\begin{align} \label{QBaction}
        S& \equiv \int_0^T dt [1+L_S+L_C]\\
        L_S & \equiv \mathrm{Tr}\left[F(t) \left(i\frac{dU(t)}{dt}U(t)^\dagger- H(t)\right)\right]\\
        L_C &\equiv \sum_j\lambda_j(t) f_j(H,s_j)
\end{align}
where both $F(t)$ (which is a matrix/operator) and $\{\lambda_j(t)\}$ (which are scalars) are Lagrange multipliers. Since $H(t)$ is Hermitian, we can assume $F(t)$ is Hermitian and $\{\lambda_j(t)\}$ are real without loss of generality.

If $T$ is minimized by a time optimal Hamiltonian $H(t)$, according to the variational principle, the variations of $S$ with respect to $H(t)$, $U(t)$, $F(t)$, $\{\lambda_j(t)\}$, $\{s_j(t)\}$, and $T$ should all vanish. In particular, the variation of $S$ with respect to $H(t)$ leads to the equation that defines $F(t)$:
\begin{equation} \label{QBF}
        F(t)=\sum_j \lambda_j(t) \frac{\partial f_j(H,s_j)}{\partial H}.
\end{equation}
The variation of $S$ with respect to $U(t)$ leads to the so-called QB equation:
\begin{equation} \label{QBequation}
    i\frac{dF(t)}{dt}=[F(t),H(t)].
\end{equation}
The variation of $S$ with respect to $T$ and $U(T)$ leads to a normalization condition:
\begin{equation}\label{QBnorm}
    \mathrm{Tr}[F(t)H(t)]=1.
\end{equation}
The variation of $S$ with respect to $\lambda_j(t)$ gives the constraint equation:
\begin{equation}\label{QBcons}
    f_j(H,s_j)=0.
\end{equation}
And finally, the variation of $S$ with respect to the slack variables $s_j(t)$ leads to:
\begin{equation}\label{QBslack}
    \lambda_j(t)\frac{\partial f_j(H,s_j)}{\partial s_j}=0.
\end{equation}

This last equation is our key addition to the standard QB method. Let us try to understand intuitively what it means. According to Eq.\,\eqref{QBslack}, we need to have either $\lambda_j(t)=0$ or $\frac{\partial f^j(H,s_j)}{\partial s_j}=0$. Notably, the case where $\lambda_j(t)=0$ is equivalent to removing the constraint $f_j(H,s_j)$ from the action, while the other possible case where $\frac{\partial f^j(H,s_j)}{\partial s_j}=0$ leads to $s_j(t)=0$. As an example, we consider the inequality constraint $|J_{1N}(t)|\le J_0$, which is captured by $f(H,s)\equiv|J_{1N}(t)|^2+|s(t)|^2 - |J_0|^2=0$. It is easy to see that $\frac{\partial f(H,s)}{\partial s}=0$ leads to $s(t)=0$, and we end up with an equality constraint given by $|J_{1N}|=J_0$. On the other hand, when the slack variable $s(t)\ne0$, we end up with $|J_{1N}(t)|<J_0$, and we treat $|J_{1N}(t)|$ as effectively unconstrained while solving the QB equation and enforce the inequality only after finding the relevant solution. 

However, because both $\lambda_j(t)$ and $s_j(t)$ are time-dependent, it appears that we have to consider the above two cases for every value of time $t$ for Eq.\,\eqref{QBslack}. Fortunately, we can avoid this complication by exploiting the fact that all variables in the action $S$ are smooth functions of $t$ as long as the  Hamiltonian $H(t)$ is a smooth function of $t$, which is indeed true for any physical (experimentally realizable) Hamiltonian. As a result, at least one of the following two conditions will always hold: (1) $\lambda_j(t)=0$ for all $t$, and (2) $s_j(t)=0$ for all $t$. It's easy to check if neither of the conditions hold, at least one of $\lambda_j(t)$ and $s_j(t)$ has to be discontinuous at some value of $t$. To avoid double counting the case where $\lambda_j(t)=s_j(t)=0$ for all $t$, we will implicitly assume that $s_j(t)$ is not always zero for case (1). As we show below for our concrete QB problem defined in Section I, this continuity analysis is crucial to making the inequality constraints manageable.

\section{Finding the fastest QST Hamiltonian}\label{TightBound}

With the aforementioned recipe for treating inequality constraints in place, our next step is to solve Eqs.\,\eqref{QBF}-\eqref{QBslack} together with the boundary condition that $U(T)$ fulfills the QST. The solution obtained will form the necessary conditions for $H(t)$ to be time optimal. We note that this will lead us to obtain a family $\mathcal{H}$ of $H(t)$ that are all at least locally time optimal Hamiltonians, and we need to compare all members of $\mathcal{H}$ to find the globally time optimal Hamiltonian among them in order to identify the speed limit of QST.

To do so, let us further simplify our effective Hamiltonian $H_{\text{eff}}$ by going into an interaction picture of its diagonal part (which only add phases to the basis states)
\begin{align}\label{InteractionHam}
    H_{\text{eff}}^I(t)&=
\begin{pmatrix}
0& \tilde{J}_{1\mathcal{A}}(t)&\tilde{J}_{1 N}(t)\\
\tilde{J}^*_{1\mathcal{A}}(t)& 0 & \tilde{J}_{\mathcal{A} N}(t)\\
\tilde{J}^*_{1 N}(t)& \tilde{J}^*_{\mathcal{A}N}(t)& 0
\end{pmatrix},
\end{align}
with the constraints:
\begin{align}\label{Iconstraints}
|\tilde{J}_{1\mathcal{A}}(t)|&=|J_{1\mathcal{A}}(t)|\le \sqrt{N-2}J_0\\
|\tilde{J}_{\mathcal{A}N}(t)|&=|J_{\mathcal{A}N}(t)|\le \sqrt{N-2}J_0\\
|\tilde{J}_{1N}(t)|&=|J_{1N}(t)|\le J_0
\end{align}

Since $H_{\text{eff}}^I(t)$ has zero diagonals, we need to add two more constraints
\begin{align}
\mathrm{Tr}\left[H(t) 
\begin{pmatrix}
1&0&0\\
0&0&0\\
0&0&-1
\end{pmatrix}\right]&=0\\
\mathrm{Tr}\left[H(t) 
\begin{pmatrix}
1&0&0\\
0&-2&0\\
0&0&1
\end{pmatrix}\right]&=0 \label{HamiltonianConstraints}
\end{align}
while a third constraint of $\mathrm{Tr}[H(t)]=0$ is not needed as removing the constraint only contributes to a global phase \cite{Carlini2006}.

Our goal now is to find the time optimal Hamiltonian $H_{\text{eff}}^I(t)$ that fulfills the QST, which requires evolving the state $|\phi_1\rangle$ to $|\phi_3\rangle$ [defined in Eq.\eqref{3basis}] up to a phase. The evolution operator in the interaction picture thus needs to take the following form at time $T$ (end of the state transfer):
\begin{equation}\label{UnitaryThree}
U(T)=\begin{pmatrix}
0&0&e^{i\phi}\\
\cos \theta e^{-i \alpha}&-\sin \theta e^{-i\beta}&0\\
\sin \theta e^{i\beta}&\cos \theta e^{i \alpha}&0
\end{pmatrix}
\end{equation}
where $\theta,\alpha,\beta$ are arbitrary angles. The phase $\phi$ is not important as one can apply a single-qubit gate (which is assumed to take negligible time) on qubit $N$ at the end of the evolution to make sure the phase acquired by $|\phi_0\rangle$ matches the phase acquired by $|\phi_1\rangle$ (which evolves to $|\phi_3\rangle$).

Introducing five Lagrange multipliers $\lambda_{1\mathcal{A}}$, $\lambda_{\mathcal{A}N}$, $\lambda_{1N}$, $\lambda_1$, and $\lambda_2$ for the above five constraint equations
[Eqs.\,\eqref{Iconstraints}-\eqref{HamiltonianConstraints}], respectively, the $F(t)$ operator in Eq.\,\eqref{QBF} takes the form:
\begin{align}\label{FThree}
F(t)&=
\begin{pmatrix}         
    \lambda_1+\lambda_2&&\lambda_{1\mathcal{A}} J_{1\mathcal{A}}&&\lambda_{1N} J_{1N}\\    \lambda_{1\mathcal{A}} J_{1\mathcal{A}}^*&&-2\lambda_2&&\lambda_{\mathcal{A}N} J_{\mathcal{A}N}\\
    \lambda_{1N} J_{1N}^*&&\lambda_{\mathcal{A}N} J_{\mathcal{A}N}^*&&-\lambda_1+\lambda_2
    \end{pmatrix}
\end{align}
where to simplify the notation we suppress the time-dependence of all terms in Eq.\,\eqref{FThree} and exchange $\tilde{J}_{1N}$, $\tilde{J}_{1\mathcal{A}}$, and $\tilde{J}_{\mathcal{A}N}$ for $J_{1N}$, $J_{1\mathcal{A}}$, and $J_{\mathcal{A}N}$ as they obey the same constraints.

The normalization condition Eq.\,\eqref{QBnorm} now reduces to:
\begin{equation}\label{ThreeNorm}
\lambda_{1\mathcal{A}}|J_{1\mathcal{A}}|^2+\lambda_{\mathcal{A}N}|J_{\mathcal{A}N}|^2+\lambda_{1N}|J_{1N}|^2=1
\end{equation}
And to deal with the inequality constraints, we require from Eq.\,\eqref{QBslack} that
\begin{align}\label{CaseConditions}
        \lambda_{1\mathcal{A}}(t)s_{1\mathcal{A}}(t)&=0 \nonumber \\ 
        \lambda_{\mathcal{A}N}(t)s_{\mathcal{A}N}(t)&=0 \nonumber \\
        \lambda_{1N}(t)s_{1N}(t)&=0
\end{align}

As we discussed towards the end of Section \ref{QBSection}, each equation above requires us to investigate two different cases, resulting in the 8 different cases listed in Table 1. For each case, we shall solve the QB equation [Eq.\,\eqref{QBequation}] together with the above Eqs.\,\eqref{UnitaryThree}-\eqref{ThreeNorm}. We defer the detailed solutions of each case to Appendix A, and provide a summary of them in Table I.

\begin{table}[h]\label{QBResultsTable}
\begin{tabular}{|c|c|c|}
\hline
~~\#~~&Case&Minimum QST Time $T$\\
\hline
~~1~~&$\lambda_{1\mathcal{A}}=\lambda_{\mathcal{A}N}=\lambda_{1N}=0$ &No minimum\\
\hline
~~2~~&$s_{1\mathcal{A}} = \lambda_{\mathcal{A}N}=\lambda_{1N}=0$&No minimum\\
\hline
~~3~~&$s_{\mathcal{A}N} = \lambda_{1\mathcal{A}}=\lambda_{1N}=0$&No minimum\\
\hline
~~4~~&$s_{1\mathcal{A}} = s_{1N} = \lambda_{\mathcal{A}N} =0$ &No minimum\\
\hline
~~5~~&$s_{\mathcal{A}N}=s_{1N}= \lambda_{1\mathcal{A}}=0$ &No minimum\\
\hline
~~6~~&$s_{1N}= \lambda_{1\mathcal{A}}=\lambda_{\mathcal{A}N}=0$ &$\pi/(2J_0)$\\
\hline
~~7~~&$s_{1\mathcal{A}} = s_{\mathcal{A}N}= \lambda_{1N} =0$ &$ \pi/\sqrt{2(N-2)J_0^2+4|\overline{J}_{1N}|^2}$\\
\hline
~~8~~&$s_{1\mathcal{A}} = s_{\mathcal{A}N}= s_{1N} = 0$ &$\pi/(J_0\sqrt{2N})$\\
\hline
\end{tabular}
\caption{Summary of the solutions to the QB equation [Eq.\,\eqref{QBequation}] and accompanying Eqs.\,\eqref{UnitaryThree}-\eqref{ThreeNorm} for 8 different cases resulting from the 3 inequality constraints in Eq.\,\eqref{CaseConditions}. The cases where no minimum time exists is due to an empty set of solutions to these equations. If the solution set is non-empty, we analytically solve the Hamiltonians obeying such solutions and find the minimum QST time. In Case 7, $\overline{J}_{1N}\equiv \frac{1}{T}\int_0^T J_{1N}(t)dt$.}
\end{table}

In each case of Table 1, whenever a member of the set of slack variables $\{s_{1\mathcal{A}}(t),s_{\mathcal{A}N}(t),s_{1N}(t)\}$ is set to zero for all time, the corresponding interaction strength $\{J_{1\mathcal{A}}(t),J_{\mathcal{A}N}(t),J_{1N}(t)\}$ is maximized for all time. On the other hand, if a member of the set of Lagrange multipliers $\{\lambda_{1\mathcal{A}}(t),\lambda_{\mathcal{A}N}(t),\lambda_{1N}(t)\}$ is set to zero for all time, then the corresponding interaction strength cannot be maximized at all time (otherwise it reduces to the previous case). And since each interaction strength must vary smoothly in time, the interaction strength can only be maximized at a finite number (including zero) of specific time points in this case. 

We provide some intuitive explanations of the results in Table 1: Case 1 will not have a minimum in the QST time because all interaction strengths $\{J_{1\mathcal{A}}(t),J_{\mathcal{A}N}(t),J_{1N}(t)\}$ are not maximized for at least some finite period of time. It is therefore always possible to speed up the QST by maximizing at least one of $\{J_{1\mathcal{A}}(t),J_{\mathcal{A}N}(t),J_{1N}(t)\}$ while increasing the other two proportionally for a finite period of time. Cases 2-5 share a common structure that the interaction strengths are not mirror symmetric, i.e. $J_{1\mathcal{A}}(t)\ne J_{A\mathcal{N}}(t)$ for some finite period of time. It makes intuitive sense that a non-mirror-symmetric Hamiltonian is likely not time optimal, although the actual proof of this is much more involved (see Appendix A). Case 6 in fact requires $J_{1\mathcal{A}}=J_{\mathcal{A}N}=0$ after solving the QB equation and therefore the effective Hamiltonian reduces to a two-level Hamiltonian with a direct coupling strength of $|J_{1N}(t)|=J_0$ between basis states $|\phi_1\rangle$ and $|\phi_3\rangle$, where it is easy to show that the minimum QST time is simply given by $\pi/(2J_0)$. The optimal Hamiltonians in Case 7 and Case 8 are similar, with the only difference being in Case 7 that one cannot maximize $|J_{1N}(t)|$ for all time. Case 7 reduces to Case 8 upon setting $|J_{1N}(t)|=J_0$.

Case 8 offers the fastest QST protocol of all cases, and we can in fact construct a simple time-independent Hamiltonian $H_{\text{eff}}^{\text{opt}}$ in the Schrodinger picture that satisfies Case 8:
\begin{equation}\label{Hopt}
H_{\text{eff}}^{\text{opt}}=
\begin{pmatrix}
0 & \sqrt{N-2} J_0 & J_0 \\
\sqrt{N-2} J_0 & -3 J_0 & \sqrt{N-2} J_0\\
J_0 & \sqrt{N-2} J_0 & 0
\end{pmatrix}
\end{equation}

Note that there are other equivalently time optimal Hamiltonians satisfying Case 8, such as $-H_{\text{eff}}^{\text{opt}}$. A family of spin Hamiltonians in the form of Eq.\,\eqref{SpinHam} can be mapped to $H_{\text{eff}}^{\text{opt}}$. One simple example is:

\begin{align}\label{DivergentFieldHamiltonian}
    H_{\text{opt}}= J_0 \left[\sum_{i\neq j}(\sigma_i^{+}\sigma_j^{-}+\mathrm{h.c.})-\frac{N}{2}(\sigma_1^z + \sigma_N^z)\right]
\end{align}

This Hamiltonian can be readily implemented in many experimental platforms with uniform, all-to-all interactions, such as trapped ions \cite{MonroeSimulation}, neutral atoms in an optical lattice coupled to a cavity \cite{Park2022}, circuit QED systems \cite{Strauch2008}, or atoms trapped near a photonic waveguide \cite{Douglas2015}. Individual addressing of the qubits is required as one needs to apply a different field along $z$ on the source and target qubits. This Hamiltonian was also previously found in the study of perfect QST in fully-connected graphs \cite{CASACCINO2009}.

For experimental realization of the time-optimal QST Hamiltonian in Eq.\,\eqref{DivergentFieldHamiltonian}, one potential concern is that the experimental Hamiltonian is usually subject to noise or imperfect control. If the fidelity of the QST deteriorates with an increasing number of qubits, then the speedup of the QST brought by additional qubits may not be useful in practice. Fortunately, as shown in Appendix \ref{NoiseAppendix}, we find that noise on the interaction and/or the field strengths in Eq.\,\eqref{DivergentFieldHamiltonian} actually have vanishing effects on the fidelity of QST as $N$ increases, assuming that the noise has a finite amplitude. This means high-fidelity, time-optimal QST can be practically achieved for a large number of qubits.

A drawback of the above optimal spin Hamiltonian $H_{\text{opt}}$ is that the field strength diverges in $N$, which makes it harder to control the field strength accurately for a larger $N$. Here we also construct an alternative spin Hamiltonian that maps to the same optimal effective Hamiltonian $H_{\text{eff}}^{\text{opt}}$ but has a constant field strength applied to the source and target qubits:

\begin{align}\label{SmallFieldHamiltonian}
    H_{\text{opt}}^{\prime}  = & J_0 \left[\sigma_1^+\sigma_N^- + \left(\sum_{i=2}^{N-1}\sigma_i^{+}\right)(\sigma_1^{-}+\sigma_N^{-})+\mathrm{h.c.} \right]\\
    &-\frac{3}{2}J_0 (\sigma_1^z + \sigma_N^z)\nonumber
\end{align}

The main drawback of this equally optimal Hamiltonian is that one needs to remove the uniform couplings among the qubits $2,3,\cdots,N-1$, which is more challenging to achieve experimentally. Nonetheless, we anticipate that noise on the interaction and/or the field strengths in Eq.\,\eqref{SmallFieldHamiltonian} will have vanishing effects on the fidelity of QST as $N$ increases similarly to in the case of noise in Eq.\,\eqref{DivergentFieldFidelity} due to the similarity between the two protocols. To our best knowledge, this Hamiltonian has not been previously found.

At $T=\pi/(J_0\sqrt{2N})$, one can verify that $e^{-i H_{\text{opt}} T}|\phi_0\rangle \propto|\phi_0\rangle$ and $e^{-i  H_{\text{opt}} T}|\phi_1\rangle \propto |\phi_3\rangle$, with some phases that only depend on $N$. The same applies to $H_{\text{opt}}^{\prime}$. The desired QST described by Eq.\,\eqref{QST} with $|\psi_{\mathcal{A}}\rangle=|0\rangle ^{\otimes {N-2}}$ can then be achieved after performing a single-qubit gate (assumed to take negligible time) on the target qubit to cancel the unwanted phases.

\section{Conclusion and Outlook}\label{Conclusion}

In this paper, we aimed to find the globally time optimal Hamiltonian for realizing QST in a subclass of Hamiltonians relevant for a fully-connected computer. By developing a new QB method that can deal with inequality constraints, we have been able to find the exact minimum time needed for QST to be implemented, with an additional assumption on the permutation symmetry of the Hamiltonian over the intermediate qubits. The QB method also allows us to construct a family of explicit, time-independent Hamiltonians that can be experimentally realized to achieve the exact speed limit of QST. This is in contrast to existing QST protocols for long-range interacting systems \cite{EldredgeStateTransfer,GuoScrambling,TranHeirarchy,TranStateTransfer} that are only qualitatively optimal.

Concretely, we recall that for spin-systems where the physical interactions decay as $1/r^\alpha$ with $r$ being the inter-particle distance, qualitatively optimal bounds on the time to perform QST have only been discovered for the case of $\alpha>D$ with protocols existing that perform QST in a time that scales as $O(\log{r})$ for $D<\alpha<2D$  \cite{TranStateTransfer}. However, in the cases of $0<\alpha<D$, the best known lower bound on the time to perform QST scales as $O(\log{(N)}/N^{1-\frac{\alpha}{D}})$ and is conjectured to be loose. Our work demonstrates that this best known Lieb-Robinson bound is certainly loose with respect to the important subclass of Hamiltonians considered here, and we conjecture that a qualitatively optimal bound for $0<\alpha<D$ should scale as $1/N^{\frac{1}{2}-\frac{\alpha}{D}}$. Our results also improve the speed of the fastest known QST protocols \cite{GuoScrambling} for $0<\alpha<D$ by a factor of at least $\sqrt{2}$. For noisy intermediate scale quantum computers, even a constant factor of speedup can noticeably improve the overall fidelity of the quantum circuit \cite{Preskill_2018}. Important to this point, we have also shown that our time-optimal QST Hamiltonian is very robust to control noises. In fact, noises of finite amplitude have vanishing effects on the fidelity of QST as the number of qubit increases.

In the future, we expect the QB method we develop here, with its ability to intuitively deal with inequality constraints, to be helpful in many other quantum optimal control problems where the interaction/drive strengths are only upper bounded. For a Hamiltonian with $M$ inequality constraints, we need to deal with $2^M$ cases. But as in the problem we studied, if certain symmetries are enforced, the number of independent constraints can be drastically reduced.

An immediate open question following this work is whether one can relax the permutation symmetry assumption we enforced on the Hamiltonian. We expect a positive answer based on how the spatial inversion (mirror) symmetry naturally emerges in the time optimal Hamiltonian from our solutions to the relevant QB equations (i.e. the cases in Table I violating the mirror symmetry all lead to no time optimal Hamiltonians). The permutation symmetry would likely require a much more cumbersome calculation. One possible way we are currently investigating to carry out such a calculation is to develop a software algorithm that can solve the QB equations automatically. Such an algorithm will also allow solutions to a variety of quantum optimal control problems for much larger systems.

Another future direction related to the current work is to find the speed limit of unconditional QST protocols or two-qubit gates for a fully-connected quantum computer or more generally, strongly long-range interacting systems. We expect our generalized QB method to be useful in this endeavor at least for small system sizes, and finding time optimal two-qubit gates is important for NISQ quantum computers even if only a small number of qubits are interacting with each other \cite{Howard2022}. The alternative approach to finding such a speed limit is through the derivations of optimal Lieb-Robinson-type bounds using the Frobenius norm instead of the operator norm, and such bounds can also be used to determine the speed limit of quantum information scrambling \cite{LucasFrobenius,Yin2020}. This approach is likely capable of providing the correct scaling of the speed limits and can be complemented by the QB method-based approach we develop here to find more accurate speed limits. 

Finally, it is interesting to see whether QB or other methods in quantum optimal theory can be applied to bound the speed limits for multi-qubit gates or the generation of many-body entangled states. Alternatively, it is also worth exploring whether the lack of scalability in most quantum optimal control methods could be compensated by the recently developed Lieb-Robinson-type bounds which often apply to thermodynamic systems but lack the precision needed for finite quantum systems.

\begin{acknowledgments}
We acknowledge funding support from the NSF RAISE-TAQS program under Grant No. CCF-1839232 and the W. M. Keck Foundation.
\end{acknowledgments}

\appendix

\section{Detailed solutions for all cases in Table 1}\label{QBAppendix}

In this appendix, we provide the details of how we solve the QB equation [Eq.\,\eqref{QBequation}] together with the Eqs.\,\eqref{UnitaryThree}-\eqref{ThreeNorm} for all cases listed in Table I. 

Defining $N_{\mathcal{A}}\equiv N-2$ and taking $|J_{1\mathcal{A}}(t)|\le J_0$ and $|J_{\mathcal{A}N}(t)|\le J_0$, we note that Eq.\,\eqref{QBequation} reduces to:
\begin{align}\label{ThreeQBAppend}
        \frac{d\lambda_1}{dt}&=\frac{d\lambda_2}{dt}=0\\
        i\frac{d(\lambda_{1\mathcal{A}}J_{1\mathcal{A}})}{dt}&=(-\lambda_1+3\lambda_2)J_{1\mathcal{A}}+(-\lambda_{1N}+\lambda_{\mathcal{A}N})J_{1N}J_{\mathcal{A}N}^*\nonumber\\
        i\frac{d(\lambda_{\mathcal{A}N}J_{\mathcal{A}N})}{dt}&=(-\lambda_1+3\lambda_2)J_{\mathcal{A}N}+(-\lambda_{1\mathcal{A}}+\lambda_{1N})J_{1N}J_{1\mathcal{A}}^*\nonumber\\
        i\frac{d(\lambda_{1N}J_{1N})}{dt}&=-2\lambda_1J_{1N}-N_{\mathcal{A}}(\lambda_{1\mathcal{A}}-\lambda_{\mathcal{A}N})J_{1\mathcal{A}}J_{\mathcal{A}N}\nonumber
\end{align}

The boundary condition Eq.\,\eqref{UnitaryThree} becomes:

\begin{align}\label{ThreeBoundaryAppend}
        \lambda_1&=0\\
        6\lambda_2\cos^2\theta&=\sqrt{N_\mathcal{A}}|\lambda_{\mathcal{A}N}J_{\mathcal{A}N}|\sin2\theta\cos(\alpha+\beta-\text{arg}[J_{\mathcal{A}N}])\nonumber\\
        \lambda_{1\mathcal{A}}J_{1\mathcal{A}}&=\lambda_{\mathcal{A}N}(J_{\mathcal{A}N} \cos^2\theta e^{ i 2\alpha}-J_{\mathcal{A}N}^* \sin^2\theta e^{-i 2\beta})\nonumber\\
        &+3 \frac{\lambda_2}{\sqrt{N_\mathcal{A}}} \sin \theta\cos\theta e^{i(\alpha-\beta)}\nonumber\\
        \lambda_{\mathcal{A}N}J_{\mathcal{A}N}&=e^{-i\phi}(\frac{\lambda_{1N}}{\sqrt{N_\mathcal{A}}}J_{1N}^*\cos\theta e^{-i\alpha}-\lambda_{1\mathcal{A}}J_{1\mathcal{A}}^*\sin\theta e^{i \beta})\nonumber\\
        \frac{\lambda_{1N}}{\sqrt{N_\mathcal{A}}}J_{1N}&=e^{-i \phi} (\frac{\lambda_{1N}}{\sqrt{N_\mathcal{A}}}J_{1N}^*\sin\theta e^{-i\beta}+\lambda_{1\mathcal{A}}J_{1\mathcal{A}}^*\cos\theta e^{i \alpha})\nonumber
\end{align}

To proceed, we will look at each of the 8 cases listed in Table I one after the other.

\subsection{$\lambda_{1\mathcal{A}}=\lambda_{\mathcal{A}N}=\lambda_{1N}=0$}\label{Case1}
This case directly contradicts with Eq.\,\eqref{ThreeNorm} and thus will not lead to any locally time optimal Hamiltonian.

\subsection{$s_{1\mathcal{A}}(t)=\lambda_{\mathcal{A}N}(t)=\lambda_{1N}(t)=0$}\label{Case2}

In this case, we replace $J_{1\mathcal{A}}(t)$ with $J_0e^{-i \phi_{1\mathcal{A}}(t)}$. The
Eqs.\,\eqref{ThreeQBAppend} and \eqref{ThreeBoundaryAppend} now become:
\begin{align}\label{Case2equations}
    1&=2N_{\mathcal{A}}\lambda_{1\mathcal{A}}J_0\\
    \lambda_2 \cos^2\theta\nonumber&=0\\
    \lambda_{1\mathcal{A}}J_0 e^{-i \phi_{1\mathcal{A}}}&=\frac{3 \lambda_2 \sin 2\theta e^{i(\alpha-\beta)}}{2\sqrt{N_{\mathcal{A}}}}\nonumber\\
    0&=\sqrt{N_{\mathcal{A}}}\lambda_{1\mathcal{A}}J_0\sin\theta e^{-i(\phi-\beta-\phi_{1\mathcal{A}})}\nonumber\\
    0&=\sqrt{N_{\mathcal{A}}}\lambda_{1\mathcal{A}}J_0\cos\theta e^{-i(\phi-\alpha-\phi_{1\mathcal{A}})}\nonumber
\end{align}

These equations are self-contradictory and again indicate that there is no locally time optimal Hamiltonian in this case.

\subsection{$s_{\mathcal{A}N}(t)=\lambda_{1\mathcal{A}}(t)=\lambda_{1N}(t)=0$}\label{Case3}

This case is very similar to Case 2. Replacing $J_{\mathcal{A}N}(t)$ with $J_0e^{-i \phi_{\mathcal{A}N}(t)}$,
Eqs.\,\eqref{ThreeQBAppend} and \eqref{ThreeBoundaryAppend} become:
\begin{align}\label{Case3equations}
    1&=2N_{\mathcal{A}}\lambda_{\mathcal{A}N}J_0\\
    \lambda_2 \cos^2\theta\nonumber&=\sqrt{N_{\mathcal{A}}}\lambda_{\mathcal{A}N}J_0\sin2\theta\cos(\alpha+\beta-\phi_{\mathcal{A}N})\\
    0&=\lambda_{\mathcal{A}N}J_0( \cos^2\theta e^{ i (2\alpha-\phi_{\mathcal{A}N})}-\sin^2\theta e^{-i (2\beta-\phi_{\mathcal{A}N})})\nonumber\\
        &+\frac{3 \lambda_2 \sin \theta\cos\theta e^{i(\alpha-\beta)}}{\sqrt{N_{\mathcal{A}}}}\nonumber\\
        0&=\sqrt{N_{\mathcal{A}}}\lambda_{\mathcal{A}N}J_0e^{-i\phi_{\mathcal{A}N}}\nonumber
\end{align}

These equations are again self-contradictory and indicate that there is no locally time optimal Hamiltonian in this case.

\subsection{$s_{1\mathcal{A}}(t)=s_{1N}(t)=\lambda_{\mathcal{A}N}(t)=0$}\label{Case4}

Replacing $J_{1\mathcal{A}}(t)$ with $J_0 e^{-i\phi_{1\mathcal{A}}(t)}$ and $J_{1N}(t)$ with $J_0 e^{-i\phi_{1N}(t)}$ in Eq.\,\eqref{ThreeQBAppend}, the equation reduces to:
\begin{align}\label{Case4QB}
    \frac{d\lambda_{1\mathcal{A}}}{dt}&=\lambda_{1N}|J_{\mathcal{A}N}|\sin(\phi_{1N}-\phi_{1\mathcal{A}}-\text{arg}(J_{\mathcal{A}N}))\\
    \lambda_{1\mathcal{A}}\frac{d\phi_{1\mathcal{A}}}{dt}&=-3\lambda_2-\lambda_{1N}|J_{\mathcal{A}N}|\cos(\phi_{1N}-\phi_{1\mathcal{A}}-\text{arg}(J_{\mathcal{A}N}))\nonumber\\
    \frac{d\lambda_{1N}}{dt}&=-N_{\mathcal{A}}\lambda_{1\mathcal{A}}|J_{\mathcal{A}N}|\sin(\phi_{1N}-\phi_{1\mathcal{A}}-\text{arg}(J_{\mathcal{A}N}))\nonumber\\
    \lambda_{1N}\frac{d\phi_{1N}}{dt}&=-N_{\mathcal{A}}\lambda_{1\mathcal{A}}|J_{\mathcal{A}N}|\cos(\phi_{1N}-\phi_{1\mathcal{A}}-\text{arg}(J_{\mathcal{A}N}))\nonumber\\
    0&=3J_{\mathcal{A}N}\lambda_2-(\lambda_{1\mathcal{A}}-\lambda_{1N})J_0^2e^{-i(\phi_{1N}-\phi_{1\mathcal{A}})}\nonumber
\end{align}

The same replacements in Eq.\,\eqref{ThreeBoundaryAppend} lead to:

\begin{align}\label{Case4Norm}
 1&=2(N_{\mathcal{A}}\lambda_{1\mathcal{A}}+\lambda_{1N})J_0^2\\
    \lambda_2\cos^2\theta&=0\nonumber\\
    \lambda_{1\mathcal{A}}J_0 e^{-i\phi_{1\mathcal{A}}}&=\frac{3\lambda_2\sin2\theta e^{i(\alpha-\beta)}}{2\sqrt{N_{\mathcal{A}}}}\nonumber\\
    0&=\lambda_{1N}\cos\theta e^{i(\phi_{1N}-\alpha)} \nonumber \\ &-\sqrt{N_{\mathcal{A}}}\lambda_{1\mathcal{A}}\sin\theta e^{i(\phi_{1\mathcal{A}}+\beta)}\nonumber\\
    \lambda_{1N}e^{-i(\phi_{1N}-\phi)}&=\sqrt{N_{\mathcal{A}}}\lambda_{1\mathcal{A}}\cos\theta e^{i(\alpha+\phi_{1\mathcal{A}})} \nonumber \\ &+\lambda_{1N}\sin\theta e^{i(\phi_{1N}-\beta)}\nonumber
\end{align}
implying that either $\lambda_2=0$ or $\cos \theta=0$ such that we will always have $\lambda_{1\mathcal{A}}(T)=0$. 

The first case implies [from Eq.\,\eqref{Case4QB}] that $\lambda_{1\mathcal{A}}=\lambda_{1N}$ which leads to a contradiction in Eq.\,\eqref{Case4Norm}. Therefore, any generated unitary of the form in Eq.\,\eqref{UnitaryThree} for this case must have $\cos\theta=0$ such that Eq.\,\eqref{Case4Norm} further reduces to: 
\begin{align}\label{Case4NormReduce}
 1&=2(N_{\mathcal{A}}\lambda_{1\mathcal{A}}+\lambda_{1N})J_0^2\\
   0&=\lambda_{1\mathcal{A}}(T)=\lambda_{1\mathcal{A}}(0)\nonumber\\
   \lambda_{1N}e^{-i\phi_{1N}}&=\pm\lambda_{1N}e^{i(\phi_{1N}-\phi-\beta)}\nonumber
\end{align}

By taking the derivative with respect to $t$ of the first equality of Eq.\,\eqref{Case4NormReduce}, we find:
\begin{equation}\label{NormTimeDerivative}
    \frac{d\lambda_{1N}}{dt}=-N_{\mathcal{A}}\frac{d\lambda_{1\mathcal{A}}}{dt}
\end{equation}

Combining Eqs.\,\eqref{Case4QB} and \eqref{NormTimeDerivative}:
\begin{align}\label{Case4Reduce}
\frac{d\lambda_{1\mathcal{A}}}{dt}&=\lambda_{1\mathcal{A}}|J_{\mathcal{A}N}|\sin(\phi_{1N}-\phi_{1\mathcal{A}}-\text{arg}(J_{\mathcal{A}N}))\\
&=\lambda_{1N}|J_{\mathcal{A}N}|\sin(\phi_{1N}-\phi_{1\mathcal{A}}-\text{arg}(J_{\mathcal{A}N}))\nonumber
\end{align}
such that Eq.\,\eqref{Case4Reduce} can be satisfied only if $\lambda_{1\mathcal{A}}=\lambda_{1N}$, $|J_{\mathcal{A}N}(t)|=0$, or $\sin(\phi_{1N}-\phi_{1\mathcal{A}}-\text{arg}(J_{\mathcal{A}N}))=0$, which implies $\frac{d\lambda_{1\mathcal{A}}}{dt}=\frac{d\lambda_{1N}}{dt}=0$. Combined with Eq.\,\eqref{Case4NormReduce}, this requires $\lambda_{1\mathcal{A}}=0$, $\lambda_{1N}\neq0$ such that Eq.\,\eqref{Case4QB} further reduces to:
\begin{align}\label{Case4ReduceFull}
        0&=\frac{d\phi_{1N}}{dt}=\sin(\phi_{1N}-\phi_{1\mathcal{A}}-\text{arg}(J_{\mathcal{A}N}))\\
        \frac{d \phi_{1\mathcal{A}}}{dt}&=-\frac{d}{dt}\text{arg}(J_{\mathcal{A}N})\nonumber\\
        \cos(n\pi)&=\cos(\phi_{1N}-\phi_{1\mathcal{A}}-\text{arg}(J_{\mathcal{A}N})),\,n\in\mathbb{Z}\nonumber\\
        3\lambda_2&=-\lambda_{1N}|J_{\mathcal{A}N}|\cos(n\pi)\nonumber\\
        |J_{\mathcal{A}N}|^2&=J_0^2\nonumber
\end{align}

Thus, Eq.\,\eqref{Case4ReduceFull} requires us to replace $J_{\mathcal{A}N}$ with $\pm J_0e^{-i\phi_{\mathcal{A}N}}$ such that we can transform the Hamiltonian of the form in Eq.\,\eqref{InteractionHam} to a basis where it is real-valued
\begin{align}\label{Case4Ham}
    H^{\prime}(t)&=e^{-i\int_0^t d\tau H_{\phi}(\tau)} (H+H_{\phi})e^{i\int_0^t d\tau H_{\phi}(\tau)}\\
    &=
    \begin{pmatrix}
    c_{1\mathcal{A}}&\sqrt{N_{\mathcal{A}}}J_0&J_0\\
    \sqrt{N_{\mathcal{A}}}J_0&0&\pm\sqrt{N_{\mathcal{A}}}J_0\\
    J_0&\pm\sqrt{N_{\mathcal{A}}}J_0&c_{1\mathcal{A}}
    \end{pmatrix}\nonumber
\end{align}
where $c_{1\mathcal{A}}=-\frac{d\phi_{1\mathcal{A}}}{dt}$ is potentially time-dependent and $H_\phi$ is given by:
\begin{align}\label{HphiForm}
H_{\phi}&=
    \begin{pmatrix}
    c_{1\mathcal{A}}&0&0\\
    0&0&0\\
    0&0&c_{1\mathcal{A}}
    \end{pmatrix}
\end{align}
Note that the above basis transformation does not affect the universality class of Eq.\,\eqref{UnitaryThree}. Then, a second iteration of the QB method can be performed on Eq.\,\eqref{Case4Ham} with respect to
\begin{equation}\label{Case4F}
    F^\prime(t)=
    \begin{pmatrix}
    \lambda_1^\prime&\sqrt{N_{\mathcal{A}}}\lambda_{1\mathcal{A}}^\prime&\lambda_{1N}^\prime\\
    \sqrt{N_{\mathcal{A}}}\lambda_{1\mathcal{A}}^\prime&0&\pm\sqrt{N_{\mathcal{A}}}\lambda_{1\mathcal{A}}^\prime\\
    \lambda_{1N}^\prime&\pm\sqrt{N_{\mathcal{A}}}\lambda_{1\mathcal{A}}^\prime&-\lambda_1^\prime
    \end{pmatrix}
\end{equation}
such that Eqs.\,\eqref{QBequation} and \eqref{QBnorm} lead to:
\begin{align}\label{Case4PrimeEq}
1&=2(2N_{\mathcal{A}}\lambda_{1\mathcal{A}}^\prime+\lambda_{1N}^\prime)J_0\\
0&=\lambda^\prime_1=\frac{d\lambda_{1\mathcal{A}}^\prime}{dt}=\frac{d\lambda_{1N}^\prime}{dt}\nonumber\\
0&=\lambda_{1\mathcal{A}}^\prime(c_{1\mathcal{A}}\mp J_0)\pm\lambda_{1N}^\prime J_0\nonumber
\end{align}

Eq.\,\eqref{Case4PrimeEq} can only be satisfied when $c_{1\mathcal{A}}$ is a constant such that the unitary  operator generated by Eq.\,\eqref{Case4Ham} can be expressed as:

\begin{align}\label{Case4Unitary}
U_{\pm}(t)&=
\begin{pmatrix}
u^{\pm}_{11}&u^{\pm}_{12}&u^{\pm}_{13}\\
u^{\pm}_{12}&u^{\pm}_{22}&\pm u^{\pm}_{12}\\
u^{\pm}_{13}&\pm u^{\pm}_{12}&u^{\pm}_{11}
\end{pmatrix}\\
u^{\pm}_{11}&=\frac{e^{it(c_{1\mathcal{A}}\pm J_0)}}{2}\nonumber\\
&+\frac{e^{i\frac{t}{2}(c_{1\mathcal{A}}\mp J_0)}}{2}(\cos\frac{t\omega_{\pm}}{2}+i\frac{c_{1\mathcal{A}}\mp J_0}{\omega_{\pm}}\sin\frac{t\omega_{\pm}}{2})\nonumber\\
u^{\pm}_{12}&=-ie^{i\frac{t}{2}(c_{1\mathcal{A}}\mp J_0)}\frac{2\Omega_{1}}{\omega_{\pm}}\sin\frac{t\omega_{\pm}}{2}\nonumber\\
u^{\pm}_{13}&=\pm(u_{11}- e^{it(c_{1\mathcal{A}}\pm J_0)})\nonumber\\
u^{\pm}_{22}&=e^{i\frac{t}{2}(c_{1\mathcal{A}}\mp J_0)}(\cos\frac{t\omega_{\pm}}{2}-i\frac{c_{1\mathcal{A}}\mp J_0}{\omega_{\pm}}\sin\frac{t\omega_{\pm}}{2})\nonumber\\
\omega_{\pm}&=\sqrt{8N_{\mathcal{A}}J^2_{0}+(c_{1\mathcal{A}}\mp J_0)^2}\nonumber
\end{align}

In order for Eq.\,\eqref{Case4Unitary} to be equivalent to a state transfer unitary operator of the form in Eq.\,\eqref{UnitaryThree}, we require that $u^{\pm}_{12}=0$ such that $t=\frac{2m\pi}{\omega_{\pm}}$ where $m\in\mathbb{N}$ in order to respect the ambiguity of the phase associated with evolving with respect to $t$. Further, we must require that $u^{\pm}_{11}=0$, such that $\frac{m(c_{1\mathcal{A}}\pm3J_0)}{\omega}=\eta$ where $\eta=2n$ when $m=\text{odd}$ and $\eta=(2n+1)$ when $m=\text{even}$ with $n\in\mathbb{Z}$. Now, to determine what value of $c_{1\mathcal{A}}\pm 3J_0$ and $m$ will minimize $t$, we will solve the related problem
\begin{align}\label{ConstantValueProblem}
\text{minimize } &g(x,y)=\frac{x}{\sqrt{q^2+y^2}}\\
\text{given } &\frac{y-r}{\sqrt{q^2+y^2}}=\gamma=\frac{z}{x},\;|z\pm x|\ge 1\;\text{and}\; x\ge 1\nonumber
\end{align}
where $q$ and $r$ are some constants. Then, we will find that 
\begin{align}\label{yValue}
y=\frac{r\pm |\gamma|\sqrt{r^2+(1-\gamma)^2q^2}}{1-\gamma^2}
\end{align}
such that the term Eq.\,\eqref{yValue} is clearly maximal when $\gamma\rightarrow 1$ which is nearest met when $\gamma=\frac{x\pm 1}{x}$. Further, without loss of generality, we can take $\gamma\ge0$ and $r\ge0$ such that 
\begin{align}\label{yMax}
y=
\begin{cases}
-\frac{rx^2+(x+1)\sqrt{r^2x^2-(2x+1)q^2}}{2x+1},&\gamma=\frac{x+1}{x}\\
-\frac{rx^2+(x-1)\sqrt{r^2x^2+(2x-1)q^2}}{2x-1},&\gamma=\frac{x-1}{x}
\end{cases}
\end{align}

We now take the derivative of $g(x,y)$ from Eq.\,\eqref{ConstantValueProblem} with respect to $x$ after substituting in the expressions of Eq.\,\eqref{yMax}.
\begin{align}\label{ConstantValueDerivative}
\frac{d g(x)}{dx}&=
\begin{cases}
\frac{-(r^3 x+rs^2 (x+2)(2x+1)-(r^2-s^2(2x+1))\tilde{r}_+)}{(2x+1)(r^2(x+1)^2+\tilde{r}_+^2+2rx(1+x)\tilde{r}_+)\tilde{r}_+\tilde{s}_+},& \gamma=\frac{x+1}{x}\\
\frac{-(r^3 x-rs^2 (x-2)(2x-1)-(r^2+s^2(2x-1))\tilde{r}_-)}{(2x-1)(r^2(x-1)^2-\tilde{r}_-^2+2rx(1+x)\tilde{r}_-)\tilde{r}_-\tilde{s}_-},& \gamma=\frac{x-1}{x}
\end{cases} \nonumber\\
\tilde{r}_+&=\sqrt{r^2x^2-s^2(2x+1)}\nonumber\\
\tilde{s}_+&=\sqrt{s^2+\frac{x^2}{(2x+1)^2}(rx+(x+1)\tilde{r}_+)^2}\nonumber\\
\tilde{r}_-&=\sqrt{r^2x^2+s^2(2x-1)}\nonumber\\
\tilde{s}_-&=\sqrt{s^2+\frac{x^2}{(2x-1)^2}(rx+(x-1)\tilde{r}_-)^2}
\end{align}
By inspection, Eq.\,\eqref{ConstantValueDerivative} is a decreasing function of $x>0$ when $\gamma=\frac{x+1}{x}$ and an increasing function of $x>0$ when $\gamma=\frac{x-1}{x}$. Given that both of these cases have $g(x)$ asymptotically approach the same value, the smallest possible value of $g(x)$ occurs for $x=1$. 

Applying the result of Eq.\,\eqref{ConstantValueProblem}, in order for Eq.\,\eqref{Case4Unitary} to be equivalent to a state transfer unitary operator of the form in Eq.\,\eqref{UnitaryThree}, we require that $c_{1\mathcal{A}}\pm3J_0=0$ and $m=1$ such that the target unitary is generated in a minimum time $T_{\text{min}}=\frac{\pi}{\sqrt{2N_{\mathcal{A}}}J_0}$. However, in this case we require $|J_{1\mathcal{A}}(t)|=J_0$ for all time, which requires $s_{1\mathcal{A}}=0$, and hence this case reduces to Case 8.

\subsection{$s_{\mathcal{A}N}(t)=s_{1N}(t)=\lambda_{1\mathcal{A}}(t)=0$}\label{Case5}
This case is very similar to Case 4. We can perform the same analysis as Case 4 by swapping the qubit labels $``1"$ and $``N"$, and we will find that locally time optimal QST can only be achieved here if $|J_{\mathcal{A}N}(t)|=J_0$ for all time. This requires $s_{\mathcal{A}N}=0$ and again reduces to Case 8.

\subsection{$s_{1N}(t)=\lambda_{1\mathcal{A}}(t)=\lambda_{\mathcal{A}N}(t)=0$}\label{Case6}

Replacing $J_{1N}(t)$ with $J_0e^{-i \phi_{1N}(t)}$,
Eqs.\,\eqref{ThreeQBAppend} and \eqref{ThreeBoundaryAppend} become:

\begin{align}\label{Case6equations}
    1&=2\lambda_{1N}J^2_{0}\\
    0&=\frac{d\phi_{1N}}{dt}=\frac{d\lambda_2}{dt}\nonumber\\
    0&=3J_{1\mathcal{A}}\lambda_2+J_{\mathcal{A}N}^*\lambda_{1N}J_0e^{-i \phi_{1N}}\nonumber\\
    0&=3J_{\mathcal{A}N}\lambda_2+J_{1\mathcal{A}}^*\lambda_{1N}J_0e^{-i \phi_{1N}}\nonumber
\end{align}

Given $\lambda_{1N}J_0\neq0$, Eq.\,\eqref{Case6equations} can be satisfied if $J_{1\mathcal{A}}=J_{\mathcal{A}N}=0$ such that $\phi_{1N}$ is an arbitrary constant, or if $\phi_{1N}=m\pi,\,m\in\mathbb{Z}$, $|J_{1\mathcal{A}}(t)|=|J_{\mathcal{A}N}(t)|$, and $\text{arg}(J_{1\mathcal{A}})=\text{arg}(J_{\mathcal{A}N})+n\pi,\,n\in\mathbb{Z}$. 

In the case where $J_{1\mathcal{A}}(t)=J_{\mathcal{A}N}(t)=0$, the unitary generated by Eq.\,\eqref{InteractionHam} can be expressed as:
\begin{equation}\label{Case6AUnitary}
U(t)=
\begin{pmatrix}
\cos J_0 t&0& -ie^{-i\phi_{1N}}\sin J_0 t\\
0&1&0\\
-ie^{-i\phi_{1N}}\sin J_0 t&0&\cos J_0t
\end{pmatrix}
\end{equation}
which is equivalent to a state transfer unitary operator of the form in Eq.\,\eqref{UnitaryThree} in a minimum time of $T_{\text{min}}=\frac{\pi}{2J_0}$.

In the case where $\phi_{1N}=m\pi,\,m\in\mathbb{Z}$, $|J_{1\mathcal{A}}(t)|=|J_{\mathcal{A}N}(t)|$, and $\text{arg}(J_{1\mathcal{A}})=\text{arg}(J_{\mathcal{A}N})+n\pi,\,n\in\mathbb{Z}$, the Hamiltonian operator of the form in Eq.\,\eqref{InteractionHam} can be transformed to a basis where it is real-valued
\begin{align}\label{Case6BHam}
    H^{\prime}(t)&=e^{-i\int_0^t d\tau H_{\phi}(\tau)} (H+H_{\phi})e^{i\int_0^t d\tau H_{\phi}(\tau)}\\
    &=
    \begin{pmatrix}
    -\frac{d}{dt}\text{arg}(J_{1\mathcal{A}})&\sqrt{N_{\mathcal{A}}}|J_{1\mathcal{A}}(t)|&\pm J_0\\
    \sqrt{N_{\mathcal{A}}}|J_{1\mathcal{A}}(t)|&0&\sqrt{N_{\mathcal{A}}}|J_{1\mathcal{A}}(t)|\\
    \pm J_0&\sqrt{N_{\mathcal{A}}}|J_{1\mathcal{A}}(t)|&-\frac{d}{dt}\text{arg}(J_{1\mathcal{A}})
    \end{pmatrix}\nonumber\\
    H_{\phi}&=
    \begin{pmatrix}
    -\frac{d}{dt}\text{arg}(J_{1\mathcal{A}})&0&0\\
    0&0&0\\
    0&0&-\frac{d}{dt}\text{arg}(J_{1\mathcal{A}})
    \end{pmatrix}\nonumber
\end{align}
where this basis transformation does not affect the constraints on the Hamiltonian. Therefore, a second iteration of the QB method can be performed on Eq.\,\eqref{Case6BHam} with respect to an $F^\prime(t)$ operator
\begin{equation}\label{Case6BF}
    F^\prime(t)=
    \begin{pmatrix}
    \lambda_1^\prime&\lambda_3^\prime&\lambda_{1N}^\prime\\
    \lambda_3^\prime&0&-\lambda_3^\prime\\
    \lambda_{1N}^\prime&-\lambda_3^\prime&-\lambda_1^\prime
    \end{pmatrix}
\end{equation}
where Eqs.\,\eqref{QBequation} and \eqref{QBnorm} become:
\begin{align}\label{Case6Bequations}
        1&=\pm2\lambda_{1N}^\prime J_0\\
        0&=\frac{d\lambda_1^\prime}{dt}=\frac{d\lambda_3^\prime}{dt}=\frac{d\lambda_{1N}^\prime}{dt}\nonumber\\
        0&=\sqrt{N_{\mathcal{A}}}|J_{1\mathcal{A}}|(\lambda_1^\prime+\lambda_{1N}^\prime)-(\frac{d}{dt}\text{arg}(J_{1\mathcal{A}})\pm J_0)\lambda_3^\prime\nonumber\\
        0&=\sqrt{N_{\mathcal{A}}}|J_{1\mathcal{A}}|(\lambda_1^\prime-\lambda_{1N}^\prime)-(\frac{d}{dt}\text{arg}(J_{1\mathcal{A}})\pm J_0)\lambda_3^\prime\nonumber\\
        0&=\sqrt{N_{\mathcal{A}}}|J_{1\mathcal{A}}|\lambda_3^\prime\pm J_0\lambda_1^\prime\nonumber
\end{align}

In order to satisfy Eq.\,\eqref{Case6Bequations}, we must require that $|J_{1\mathcal{A}}|=\lambda_1^\prime=0$. Given the phase term $\text{arg}(J_{1\mathcal{A}})$ will have no physical source, we must further require that $\frac{d}{dt}\text{arg}(J_{1\mathcal{A}})=0$. Then, the unitary generated by Eq.\,\eqref{Case6BHam} can be expressed by Eq.\,\eqref{Case6AUnitary}, which is equivalent to a state transfer unitary operator of the form in Eq.\,\eqref{UnitaryThree} in a minimum time of $T_{\text{min}}=\frac{\pi}{2J_0}$.

\subsection{$s_{1\mathcal{A}}(t)=s_{\mathcal{A}N}(t)=\lambda_{1N}(t)=0$}\label{Case7}

Replacing $J_{1\mathcal{A}}(t)$ with $J_0 e^{-i\phi_{1\mathcal{A}}(t)}$ and $J_{\mathcal{A}N}(t)$ with $J_0 e^{-i\phi_{\mathcal{A}N}(t)}$, Eqs.\,\eqref{ThreeQBAppend} and \eqref{ThreeBoundaryAppend} reduce to:

\begin{align}\label{Case7equations}
    0&=N_{\mathcal{A}}(\lambda_{1\mathcal{A}}-\lambda_{\mathcal{A}N})J^2_{0}\\
    1&=2N_{\mathcal{A}}(\lambda_{1\mathcal{A}}+\lambda_{\mathcal{A}N})J^2_{0}\nonumber\\
    \frac{d\lambda_{1\mathcal{A}}}{dt}&=-\lambda_{\mathcal{A}N}|J_{1N}|\sin(\text{arg}(J_{1N})-\phi_{1\mathcal{A}}-\phi_{\mathcal{A}N})\nonumber\\
    \lambda_{1\mathcal{A}}\frac{d\phi_{1\mathcal{A}}}{dt}&=-3\lambda_2+\lambda_{\mathcal{A}N}|J_{1N}|\cos(\text{arg}(J_{1N})-\phi_{1\mathcal{A}}-\phi_{\mathcal{A}N})\nonumber\\
    \frac{d\lambda_{\mathcal{A}N}}{dt}&=\lambda_{1\mathcal{A}}|J_{1N}|\sin(\text{arg}(J_{1N})-\phi_{1\mathcal{A}}-\phi_{\mathcal{A}N})\nonumber\\
    \lambda_{\mathcal{A}N}\frac{d\phi_{\mathcal{A}N}}{dt}&=3\lambda_2-\lambda_{1\mathcal{A}}|J_{1N}|\cos(\text{arg}(J_{1N})-\phi_{1\mathcal{A}}-\phi_{\mathcal{A}N})\nonumber
\end{align}

In order to satisfy Eq.\,\eqref{Case7equations}, we must require that $\lambda_{1\mathcal{A}}(t)=\lambda_{\mathcal{A}N}(t)$ and $\frac{d\phi_{1\mathcal{A}}}{dt}=-\frac{d\phi_{\mathcal{A}N}}{dt}$, such that it can only be satisfied if $|J_{1N}|=0$ or if $\sin(\text{arg}(J_{1N})-\phi_{1\mathcal{A}}-\phi_{\mathcal{A}N})=0$ 

In the case where $|J_{1N}(t)|=0$, we can transform the Hamiltonian of the form in Eq.\,\eqref{InteractionHam} to a basis where it is constant, real-valued
\begin{align}\label{Case7AHam}
    H^{\prime}(t)&=e^{-i t H_{\phi}} (H+H_{\phi})e^{i t H_{\phi}}\\
    &=
    \begin{pmatrix}
    c_{1\mathcal{A}}&\sqrt{N_{\mathcal{A}}}J_0&0\\
    \sqrt{N_{\mathcal{A}}}J_0&0&\sqrt{N_{\mathcal{A}}}J_0\\
    0&\sqrt{N_{\mathcal{A}}}J_0&c_{1\mathcal{A}}
    \end{pmatrix}\nonumber
\end{align}
where $c_{1\mathcal{A}}=-\frac{d\phi_{1\mathcal{A}}}{dt}=12N_{\mathcal{A}}\lambda_2J_0^2$ and $H_{\phi}$ is given in Eq.\,\eqref{HphiForm}. Given that $c_{1\mathcal{A}}$ is proportional to $\lambda_2$ it can be taken as a free parameter. Now, the unitary operator generated by Eq.\,\eqref{Case7AHam} can be expressed
\begin{align}\label{Case7AUnitary}
U(t)=&
\begin{pmatrix}
u_{11}&u_{12}&u_{13}\\
u_{12}&u_{22}&u_{12}\\
u_{13}&u_{12}&u_{11}
\end{pmatrix}\\
u_{11}&=\frac{e^{-it c_{1\mathcal{A}}}}{2}+\frac{e^{-i\frac{t}{2}c_{1\mathcal{A}}}}{2}(\cos\frac{t\omega}{2}-i\frac{c_{1\mathcal{A}}}{\omega}\sin\frac{t\omega}{2})\nonumber\\
u_{12}&=-ie^{-i\frac{t}{2}c_{1\mathcal{A}}}\frac{\sqrt{N_{\mathcal{A}}}J_0}{\omega}\sin\frac{t\omega}{2}\nonumber\\
u_{13}&=u_{11}- e^{-itc_{1\mathcal{A}}}\nonumber\\
u_{22}&=e^{-i\frac{t}{2}c_{1\mathcal{A}}}(\cos\frac{t\omega}{2}+i\frac{c_{1\mathcal{A}}}{\omega}\sin\frac{t\omega}{2})\nonumber\\
\omega&=\sqrt{8N_{\mathcal{A}}J^2_{0}+c^2_{1\mathcal{A}}}\nonumber
\end{align}
In order for Eq.\,\eqref{Case7AUnitary} to be equivalent to a state transfer unitary operator of the form in Eq.\,\eqref{UnitaryThree}, we require that $u_{12}=0$ such that $t=\frac{2m\pi}{\omega}$ where $m\in\mathbb{N}$ in order to respect the ambiguity of the phase associated with evolving with respect to $t$. Further, we must require that $u_{11}=0$, such that $\frac{mc_{1\mathcal{A}}}{\omega}=0$ where $\eta=2 n$ when $m=\text{odd}$ and $\eta=(2n+1)$ when $m=\text{even}$ with $n\in\mathbb{Z}$. 

Applying the result of Eq.\,\eqref{ConstantValueProblem}, we find that in order for Eq.\,\eqref{Case7AUnitary} to be equivalent to a state transfer unitary operator of the form in Eq.\,\eqref{UnitaryThree}, we require that $c_{1\mathcal{A}}=0$ and $m=1$ such that the target unitary is generated in a minimum time $T_{\text{min}}=\frac{\pi}{\sqrt{2(N-2)J_0^2}}$.

 In the case where $\sin(\text{arg}(J_{1N})-\phi_{1\mathcal{A}}-\phi_{\mathcal{A}N})=0$, we can transform the Hamiltonian of the form in Eq.\,\eqref{InteractionHam} to a basis where it is real-valued
\begin{align}\label{Case7BHam}
    H^{\prime}(t)&=e^{-i t H_{\phi}} (H+H_{\phi})e^{i t H_{\phi}}\\
    &=
    \begin{pmatrix}
    c_{1\mathcal{A}}-|J_{1N}|&\sqrt{N_{\mathcal{A}}}J_0&|J_{1N}|\\
    \sqrt{N_{\mathcal{A}}}J_0&0&\sqrt{N_{\mathcal{A}}}J_0\\
    |J_{1N}|&\sqrt{N_{\mathcal{A}}}J_0&c_{1\mathcal{A}}-|J_{1N}|
    \end{pmatrix}\nonumber\\
    H_{\phi}&=
    \begin{pmatrix}
    c_{1\mathcal{A}}-|J_{1N}|&0&0\\
    0&0&0\\
    0&0&c_{1\mathcal{A}}-|J_{1N}|
    \end{pmatrix}\nonumber
\end{align}
where $c_{1\mathcal{A}}=-\frac{d\phi_{1\mathcal{A}}}{dt}=12N_{\mathcal{A}}\lambda_2J_0^2$. Given that $c_{1\mathcal{A}}$ is proportional to $\lambda_2$ it can be taken as a free-parameter, and because the terms dependent on $|J_{1N}(t)|$ commute with the rest of the Hamiltonian, the unitary operator generated by Eq.\,\eqref{Case7BHam} can be expressed
\begin{align}\label{Case7BUnitary}
U(t)&=
\begin{pmatrix}
u_{11}&u_{12}&u_{13}\\
u_{12}&u_{22}&u_{12}\\
u_{13}&u_{12}&u_{11}
\end{pmatrix}\\
u_{11}&=\frac{e^{-it(c_{1\mathcal{A}}-2|\overline{J}_{1N}|)}}{2}+\frac{e^{-i\frac{t}{2}c_{1\mathcal{A}}}}{2}(\cos\frac{t\omega}{2}+i\frac{c_{1\mathcal{A}}}{\omega}\sin\frac{t\omega}{2})\nonumber\\
u_{12}&=-ie^{-i\frac{t}{2}c_{1\mathcal{A}}}\frac{\sqrt{N_{\mathcal{A}}}J_0}{\omega}\sin\frac{t\omega}{2}\nonumber\\
u_{13}&=u_{11}- e^{it(c_{1\mathcal{A}}-2|\overline{J}_{1N}|)}\nonumber\\
u_{22}&=e^{-i\frac{t}{2}c_{1\mathcal{A}}}(\cos\frac{t\omega}{2}+i\frac{c_{1\mathcal{A}}}{\omega}\sin\frac{t\omega}{2})\nonumber\\
\omega&=\sqrt{8N_{\mathcal{A}}J^2_{0}+c^2_{1\mathcal{A}}}\nonumber
\end{align}

In order for Eq.\,\eqref{Case7BUnitary} to be equivalent to a state transfer unitary operator of the form in Eq.\,\eqref{UnitaryThree}, we require that $u_{12}=0$ such that $t=\frac{2m\pi}{\omega}$ where $m\in\mathbb{N}$ in order to respect the ambiguity of the phase associated with evolving with respect to $t$. Further, we must require that $u_{11}=0$, such that $\frac{m(c_{1\mathcal{A}}-4 |\overline{J}_{1N}|)}{\omega}=0$ where $\eta=2 n$ when $m=\text{odd}$ and $\eta=(2n+1)$ when $m=\text{even}$ with $n\in\mathbb{Z}$. 

Applying the result of Eq.\,\eqref{ConstantValueProblem}, we find that in order for Eq.\,\eqref{Case7BUnitary} to be equivalent to a state transfer unitary operator of the form in Eq.\,\eqref{UnitaryThree}, we require that $c_{1\mathcal{A}}=4 |\overline{J}_{1N}|\le4J_0$ and $m=1$ such that the target unitary is generated in a minimum time $T_{\text{min}}=\frac{\pi}{\sqrt{2}\sqrt{N_{\mathcal{A}}J_0^2+2\overline{J}_{1N}^2}}\ge\frac{\pi}{\sqrt{2N}J_0}$ which saturates only when $|J_{1N}(t)|=J_0$. Hence, at its saturation limit, this case reduces to Case 8.

\subsection{$s_{1\mathcal{A}}(t)=s_{\mathcal{A}N}(t)=s_{1N}(t)=0$}\label{Case8}
This case can be reached by taking $|J_{1N}(t)|=J_0$ in the above Case 7, which results in $T_{\text{min}}=\frac{\pi}{\sqrt{2N}J_0}$ directly. Alternatively, we can arrive at this case from Case 4 or Case 5, which results in the same minimal QST time.

\section{Noise Analysis for time optimal QST Hamiltonians}\label{NoiseAppendix}

In Eqs.\,\eqref{DivergentFieldHamiltonian} and \eqref{SmallFieldHamiltonian}, two experimentally realizable Hamiltonians that can implement QST in the minimal time are presented. In practice, these Hamiltonians will experience some noise in their implementation. Here we provide an analysis of the performance of QST using Eq.\,\eqref{DivergentFieldHamiltonian} when the control on the interaction strengths and single-qubit fields is assumed to be noisy. We expect similar results for the QST Hamiltonian in Eq.\,\eqref{SmallFieldHamiltonian}.

We add noise terms to the Hamiltonian in Eq.\,\eqref{DivergentFieldHamiltonian} as:
\begin{equation}\label{NoisyDivergentFieldHam}
 H_{\text{noisy}} = H_{\text{opt}} + \sum_{i\neq j}\epsilon_{ij} (\sigma_i^{+}\sigma_j^{-}+\mathrm{h.c.}) + \epsilon_1\sigma_1^z + \epsilon_N \sigma_N^z   
\end{equation}
where $\{\epsilon_{ij}\}$ are noisy coefficients sampled from a normal distribution of zero mean and standard deviation $\sigma_c$, while $\{\epsilon_1,\epsilon_N\}$ are sampled from a normal distribution of zero mean and standard deviation $\sigma_f$. 

Since this noisy Hamiltonian does not change the state of the system if all qubits are in $|0\rangle$, the fidelity of the QST for an arbitrary state of the source qubit is lower bounded by the fidelity of performing QST when the source qubit is in $|1\rangle$. We can express this latter fidelity as
\begin{equation}\label{DivergentFieldFidelity}
F =\langle \phi_1 | e^{i H_{\text{opt}} t} e^{-i H_{\text{noisy}} t} | \phi_1 \rangle
\end{equation}
where $|\phi_1\rangle$ is defined in Eq.\,\eqref{3basis} and $t=\frac{\pi}{\sqrt{2N}}$.

In the following, we will show that the infidelity $1-F$ should scale linearly in $\sigma_c$ and $\sigma_f$, and more importantly, as $1/\sqrt{N}$. As a result, as long as the noise amplitudes $\sigma_c$ and $\sigma_f$ are finite, the infidelity should vanish in the large $N$ limit. Alternatively, we can argue that the infidelity can remain finite and small as long as $\sigma_c, \sigma_f \ll \sqrt{N}$. Therefore, we believe our time optimal QST protocol is robust against control noises.

We first provide some intuitive analysis to explain the above mentioned scaling. If only the noise terms $\epsilon_1\sigma_1^z + \epsilon_N \sigma_N^z$ are present in Eq.\,\eqref{NoisyDivergentFieldHam}, a simple first-order perturbation theory shows that 
\begin{equation}\label{FieldPertInfidelity}
1-F \approx  |\langle \phi_1 | (\epsilon_1 - \epsilon_N) t |\phi_1 \rangle|
\end{equation}
Since $t=O(1/\sqrt{N})$, we have $1-F = O(\sigma_f/\sqrt{N})$.

Next, we consider the case where only the noise on the interactions $\sum_{i\neq j}\epsilon_{ij} (\sigma_i^{+}\sigma_j^{-}+\mathrm{h.c.}$) are present in Eq.\,\eqref{NoisyDivergentFieldHam}. Intuitively, these additional interactions are disordered and should have little effect on the QST. The time optimal QST achieved by Eq.\,\eqref{DivergentFieldHamiltonian} is through the collective couplings between the qubits $1,N$ and the qubits $2,3,\cdots,N$. We therefore expect the noise on the interactions to mainly perturb the collective coupling strength $J_{1\mathcal{A}}$ and $J_{\mathcal{A}N}$ in Eq.\,\eqref{HamThreeDiag}, with an amplitude of $\sigma_c$. Since $J_{1\mathcal{A}}=J_{\mathcal{A}N}=O(\sqrt{N})$ for Eq.\,\eqref{DivergentFieldHamiltonian}, as long as $\sigma_c \ll \sqrt{N}$, the effective Hamiltonian in Eq.\,\eqref{HamThreeDiag} is only weakly perturbed and a small infidelity is therefore expected.

\begin{figure}[h]
\centering
\includegraphics[width=\columnwidth]{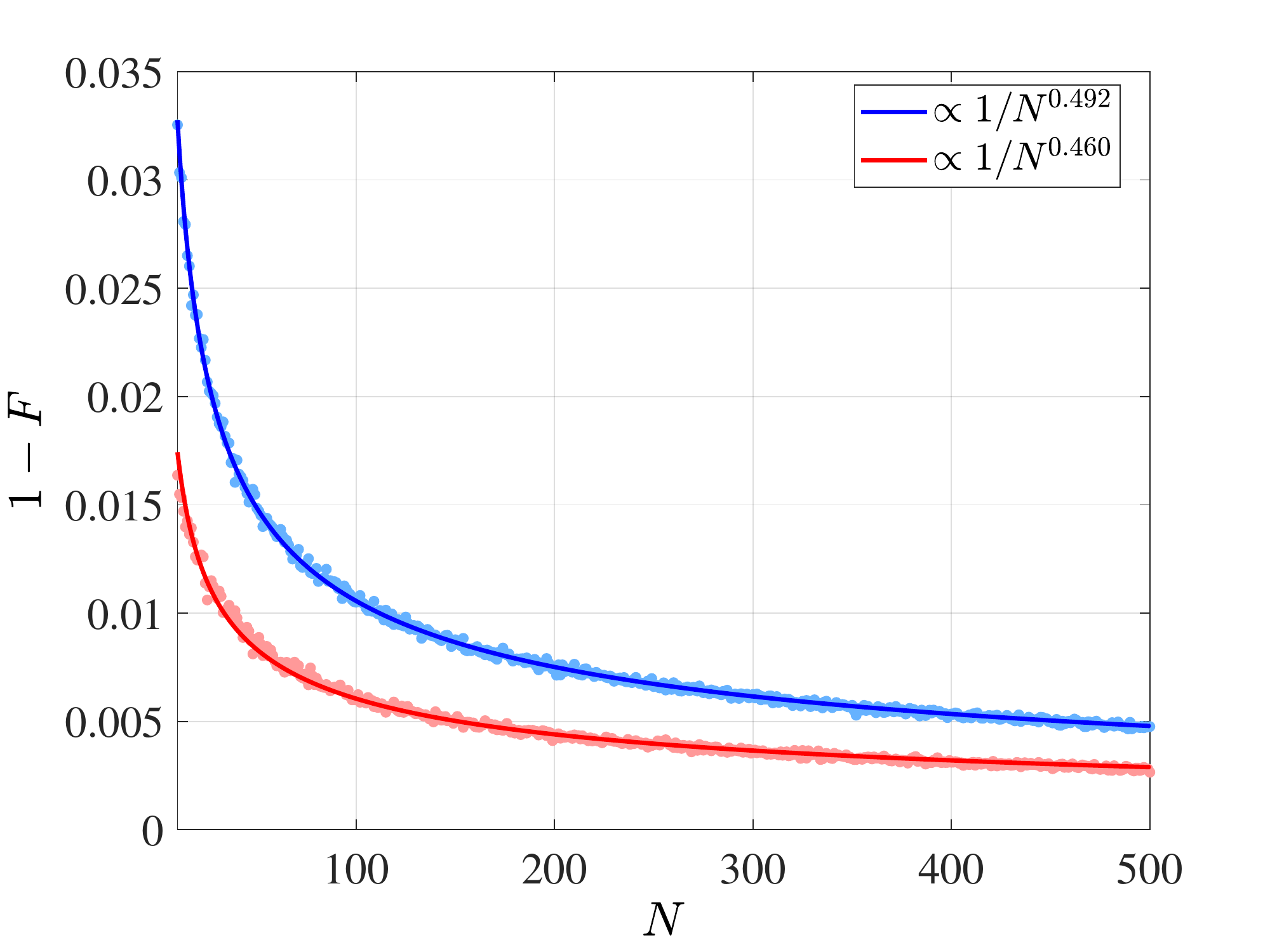}
\caption{Dependence of the QST infidelity on the number of qubits $N$ for fixed noise amplitudes. The blue dots represent $1-F$ calculated using Eq.\,\eqref{DivergentFieldFidelity} for $\sigma_c=0.1$ and $\sigma_f=0$, while the red dots correspond to the case of $\sigma_f=0.1$ and $\sigma_c=0$. The blue (red) solid line is a power-law fit to the blue (red) dots.}\label{DivergentFieldPlot}
\end{figure}

\begin{figure}[h]
\centering
\includegraphics[width=\columnwidth]{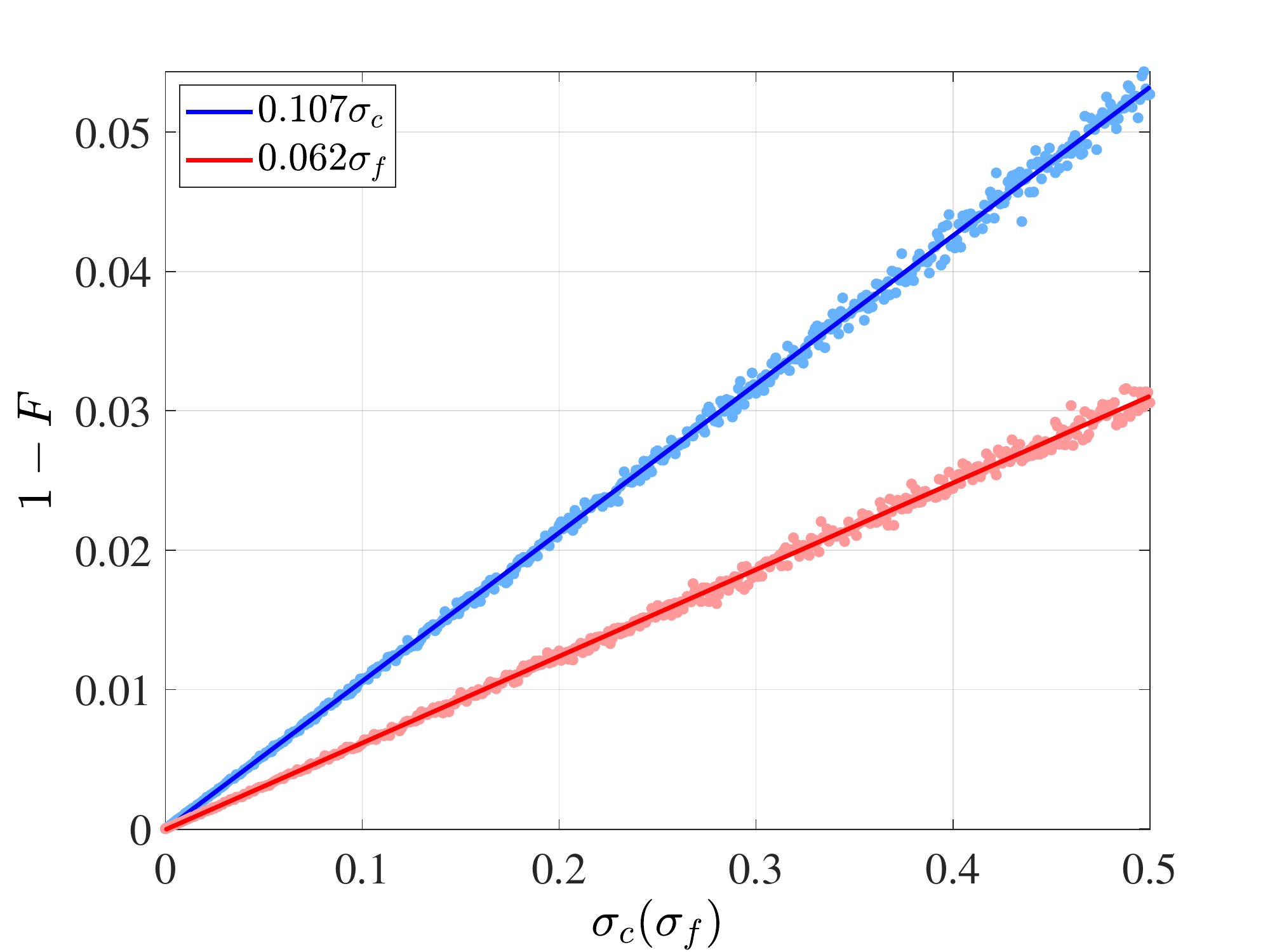}
\caption{Dependence of the QST infidelity on the noise amplitudes $\sigma_c$ and $\sigma_f$ for $N=100$ qubits. The blue dots represent $1-F$ calculated using Eq.\,\eqref{DivergentFieldFidelity} for $\sigma_f=0$ and varying $\sigma_c$, while the red dots have $\sigma_c=0$ and varying $\sigma_f$. The blue (red) solid line is a linear fit to the blue (red) dots.}\label{DivergentFieldSigmaPlot}
\end{figure}

We have also performed exact numerical calculations to support the above intuitive analysis. By averaging $1-F$ calculated using Eq.\,\eqref{DivergentFieldFidelity} over 1000 different random realizations of the noisy Hamiltonian in Eq.\,\eqref{NoisyDivergentFieldHam}, we show the dependence of this statistically averaged infidelity as a function of $N$ in Fig.\,\ref{DivergentFieldPlot} and as a function of $\sigma_c$ or $\sigma_f$ in Fig.\,\ref{DivergentFieldSigmaPlot}. We clearly see that the dependence of the infidelity on $N$ is approximately given by $O(1/\sqrt{N})$, while the infidelity scales linearly with either $\sigma_c$ or $\sigma_f$. For a system of $N=500$ qubits, the QST fidelity is above $99.5\%$ even for a fairly large noise amplitude of $\sigma_{c,f}=0.1$, i.e. $10\%$ of the interaction strength.

\pagebreak

\bibliographystyle{apsrev4-1}
\bibliography{refs}

\end{document}